\RequirePackage{fix-cm}
\documentclass[smallextended]{svjour3}       % onecolumn (second format)
\smartqed  % flush right qed marks, e.g. at end of proof
\usepackage{graphicx}
\usepackage{amssymb}
\usepackage{multirow}
\usepackage{bbding}
\begin{document}

\title{Collusive Attacks to ``Circle-Type'' Multi-party Quantum Key Agreement Protocols%\thanks{Grants or other notes
}


\author{Bin Liu         \and
        Di Xiao        \and
        Heng-Yue Jia      \and
        %Wei Huang   \and
        Run-Zong Liu
}

\institute{B. Liu(\Envelope)\and D. Xiao\and R.-Z. Liu \at
Key Laboratory of Dependable Service Computing in Cyber Physical Society (Chongqing University) of Ministry of Education,
College of Computer Science, Chongqing University, Chongqing 400044, China\\
\email{liubin31416@gmail.com.com}
\and H.-Y. Jia \at
School of Information, Central University of Finance and Economics, Beijing 100081, China\\
              %Fax: +086-13466672880\\
}
\date{Received: date / Accepted: date}

\date{Received: date / Accepted: date}
\maketitle

\begin{abstract}
We find that existing multi-party quantum key agreement (MQKA) protocols designed for fairness of the key are, in fact, unfair.
Our analysis shows that these protocols are sensitive to collusive attacks; that is, dishonest participants can collaborate to predetermine the key without being detected. In fact, the transmission structures of the quantum particles in those unfair MQKA protocols, three of which have already been analyzed, have much in common. We call these unfair MQKA protocols circle-type MQKA protocols. Likewise, the transmission structures of the quantum particles in MQKA protocols that can resist collusive attacks are also similar. We call such protocols complete-graph-type MQKA protocols. A MQKA protocol also exists that can resist the above attacks but is still not fair, and we call it the tree-type MQKA protocol.
We first point out a common, easily missed loophole that severely compromises the fairness of present circle-type MQKA protocols. Then we show that two dishonest participants at special positions can totally predetermine the key generated by circle-type MQKA protocols. We anticipate that our observations will contribute to secure and fair MQKA protocols, especially circle-type protocols.
%We find that part of the existing multi-party quantum key agreement (MQKA) protocols which pursue the fairness of the key are not totally fair indeed. Our analysis shows that they are sensitive to collusive attacks, i.e., some of the dishonest participants can collaborate to predetermine the key without being detected. In fact, the transmission structures of the quantum particles in those unfair MQKA protocols, three of which have already been analyzed, have much in common. We call these unfair MQKA protocols the circle-type MQKA protocols. And the transmission structures of the quantum particles in the rest MQKA protocols, which can resist collusive attacks, are also similar. We call such protocols the complete-graph-type MQKA protocols. Besides, there exists a MQKA protocol which can resist the above attacks but is still not fair, and we call it the tree-type MQKA protocol.
%In this paper, we firstly point out a common loophole which is easy to be neglected while severely compromises the fairness of the present circle-type MQKA protocols, and then we show that two dishonest participants at special positions can totally predetermine the key generated by the circle-type MQKA protocols.
%We hope contains of this paper would avoid future detours in pursuing secure and fair MQKA protocols, especially the circle-type ones.

\end{abstract}

\section{Introduction}
\label{intro}
Key establishment (KE) is an important cryptographical primitive, which allows participants to share a common secret key via an insecure channel. KE may be broadly subdivided into key agreement (KA) and key distribution (KD) \cite{Menezes1997}. KD is also called key transport.
In a KD protocol, one party creates a secret key, and securely distributes it to the other(s). And in a KA protocol, a shared secret is derived by two (or more) parties as a function of information contributed by each of them \cite{Menezes1997,Mitchell1998,Ateniese2000}.
%In a KA protocol, the generated key cannot be determined by any set of the participants except the universal set. More to the point, the generated key cannot be leaked to anyone else.
The main difference between a KA protocol and a KD protocol is that the key is generated by all the participants together in the former, but by one alone in the latter.
The security of the classical key agreement is based on the computation complexity. However, along with the proposing of efficient algorithms and the development of the computing capability, especially the rapid development of quantum computer, classical key agreement faces more and more austere challenges \cite{Shor,Grover}.

In the last three decades, quantum cryptography has become a hot topic in cryptography. Various of quantum cryptographic protocols have been proposed, such as quantum key distribution \cite{BB84,Liu11,Jin2013,nature14,T3,T10,Huang2011}, quantum secret sharing \cite{CGL99,HBB99,KKI99}, quantum secure direct communication \cite{LL02,BF02,GaoOC10,Huang2012}, quantum private comparison \cite{Yang2009D,Chen2009,LiuQPC12}, and so on \cite{GaoPLA07,GaoPRL08,LiuQPQ15,Huang2014}.
Quantum key agreement (QKA) uses quantum mechanics to guarantee the security and the fairness of the generated keys. Different from the security of the classical key agreement which might be susceptible to the strong ability of quantum computation, the security of QKA is simply based on physical principles such as Heisenberg uncertainty principle and quantum no-cloning theorem. Consequently, QKA can stand against the threat from an attacker with the ability of quantum computation \cite{GRTZ02,TTMTT07}. Since the first QKA protocol was proposed by Zhou et al. in 2004 \cite{Zhou}, lots of QKA protocols have been proposed, including both the two-party ones\cite{Tsai,Chong,HuangQIP14,HuangIJTP14,Shen14} and the multi-party ones \cite{QKA_Liu,MQKA_Yin,He15,Xu14,QKA2012e,MQKA_Yin2,SunvsLiu,Chitra14,Zhu15,Sun2015QIP,Sun2015IJTP,HuangvsSun}.

What draws special attention is that the security of QKA protocol is more complicated than that of QKD. The generated key in a QKA protocol should not be determined by any non-trivial subset of the participants. And this is a difficulty in the design of QKA protocols, especially the multi-party ones, which need not only to resist the attacks from the single participant as in two-party ones, but also to prevent the collusive attacks where part of the participants cooperate to cheat the other(s). %In 2013, Shi et al. proposed the first MQKA protocol \cite{QKA2012e}, however Liu et al. find that their protocol is not fair and proposed a secure one \cite{QKA_Liu}. In the same year, a MQKA protocol \cite{SunvsLiu} which is more efficient than Liu et al.'s has been presented, while Huang et al. find that they are sensitive to a collusive attack \cite{HuangvsSun}. Recently, Zhu et al. \cite{Zhu15} pointed out that a newly proposed MQKA \cite{Chitra14} protocol is not fair since two dishonest participants can conclude to determine the shared key alone.

According to the transmission structures of quantum particles, we divide all the previous multi-party QKA (MQKA) protocols into three categories (See Fig. \ref{TT} below). In the first category, every participant sends each of the other participants a sequence of particles which carries the information of his/her personal secret key \cite{QKA_Liu,MQKA_Yin,He15}, and we call them the complete-graph-type MQKA (CGT-MQKA) protocols. While in the second category, every participant only sends out one sequence, which will be operated by each of the other participants in turn and finally sent back to the one who prepared it \cite{QKA2012e,MQKA_Yin2,SunvsLiu,Chitra14,Zhu15,Sun2015QIP,Sun2015IJTP}, and we call these ones the circle-type MQKA (CT-MQKA) protocols. Obviously, CT-MQKA protocols are more efficient than CGT-MQKA ones in the sense that they consume less quantum resource. However, it is a challenge to design an unconditionally fair CT-MQKA protocol. The first MQKA protocol\cite{QKA2012e} is just a circle-type one; however it has been proved unfair,  in Ref \cite{QKA_Liu}, since any single dishonest participant can totally predetermine the key. In the meantime, the first CGT-MQKA protocol, which has been proved fair against both single attacks and collusive attacks, has been proposed \cite{QKA_Liu}. Afterwards, people continued to explore new CT-MQKA protocols because of the higher efficiency. In the later proposed CT-MQKA protocols \cite{MQKA_Yin2,SunvsLiu,Chitra14,Zhu15,Sun2015QIP,Sun2015IJTP}, none of the participants can predetermine the key without cooperating with others. While the loopholes concerning collusive attacks are unimpressive and easily been ignored in the fairness analysis. Unfortunately, we find that all the existing CT-MQKA protocols are sensitive to collusive attacks.
Besides the two categories above, another MQKA protocol \cite{Xu14} exists which we call the tree-type MQKA (TT-MQKA) protocol (according to its special transmission structure).

This paper focuses on the collusive attacks against the MQKA protocols. We prove that in some MQKA protocols, just two dishonest participants at special positions can totally predetermine the generated key. Since all the existing CT-MQKA protocols cannot resist this kind of attacks, we think that our observations will contribute to secure and fair MQKA protocols, especially circle-type protocols. The rest of this paper is organized as follows. Three categories of the MQKA protocols, i.e., the complete-graph-type one, the circle-type one, and the tree-type one, are introduced in section 2. In section 3, we mainly discuss the fairness of CT-MQKA protocols against collusive attacks.
%provide a comprehensive analysis of the existing MQKA protocols against collusive attacks, indicating which are secure against collusive attacks and which are not,
A short conclusion and a brief discussion on another attack on QKA, which is also easy to be ignored, are given in section 4. %Analysis of the existing

\section{Three Categories of MQKA Protocols}
%As we known, about 10 MQKA protocols have been proposed since 2013 \cite{QKA_Liu,He15,Xu14,QKA2012e,MQKA_Yin,MQKA_Yin2,SunvsLiu,HuangvsSun,Chitra14,Zhu15}, and

MQKA protocols are difficult to design because of the severe requirement that the generated key cannot be determined by any nontrivial set of the participants. There mainly emerged three ways to guarantee the fairness of the generated key according to the previous MQKA protocols.
The first two ways are similar in the sense that in both of them, each participant $P_i$ first generates a personal key $K_i$, then through one of the above protocols, he securely receives all the others' personal keys \cite{QKA_Liu,He15} or the result of a bitwise exclusive OR on all the others' personal keys \cite{MQKA_Yin,QKA2012e,MQKA_Yin2,SunvsLiu,Chitra14,Zhu15,Sun2015QIP,Sun2015IJTP}, and the final key is the bitwise exclusive OR result on all the personal keys. While in the protocol proposed in Ref. \cite{Xu14}, the key is generated by the random measurement results to the $N$-party GHZ sates
\begin{equation}\label{ghz}
    \frac{1}{\sqrt{2}}(|00\cdots0\rangle+|11\cdots1\rangle)_{1,2,\ldots, N}.
\end{equation}
The overall processes of these three categories of MQKA protocols can be summarized as follows.
\begin{itemize}
  \item In the first category \cite{QKA_Liu,MQKA_Yin,He15}, every participant directly sends his/her personal keys to each of the other participants. Then each participant, for example $P_i$, performs a bitwise exclusive OR on all the keys, including his/her own key, to generate the final key. The difference between these three protocols is that the one in Ref \cite{QKA_Liu} employs one-way transmission where the keys are encoded in the quantum states directly, while the rest ones \cite{MQKA_Yin,He15} employ two-way transmission where the keys are encoded in different unitary operations. And all of them employ decoys states to detect the potential attacks.
  \item In the second category \cite{QKA2012e,MQKA_Yin2,SunvsLiu,Chitra14,Zhu15,Sun2015QIP,Sun2015IJTP}, every participant ($P_i$) generates a sequence of entangled states and sends (part of) them to the others. And the sequence of the particles sent by $P_i$ is denoted as $S_i$. After other participants encoded their personal keys in $S_i$ in turn, $S_i$ would be sent back to $P_i$. Then $P_i$ measures the entangled states to get the result of the bitwise exclusive OR on all the others' personal keys, denoted as $K_{-i}$. Finally, he performs a bitwise exclusive OR on $K_i$ and $K_{-i}$ to get to final key. In most of them, $S_i$ ``runs'' a circle begin with $P_{i+1}$ and ends with $P_{i-1}$. An exception is the protocol in Ref. \cite{Sun2015IJTP}, $P_i$ generates two sequences $S_{34}$ and $S_{65}$, each of which ``runs'' half of the circle.
  \item In Ref \cite{Xu14}, one participant (Alice) generate a sequence of GHZ states as in Eq. \ref{ghz}. For each of the GHZ states, Alice delivers each of the other participants one of its particles. Some of the GHZ states are used to perform the detections, while the others are used to generate the final key. It's worth noting that in this protocol, every participant will communicate with each of the other participants in the detection processes.
\end{itemize}

\begin{figure}
  \centering
  \includegraphics[width=12cm]{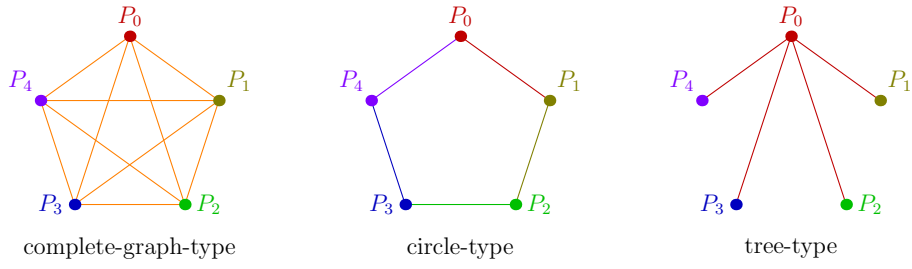}\\
  \caption{Three categories of MQKA protocols in graph, where the vertices represent the participants and the (undirected) vertices represent the transmissions of quantum states between the participants.}\label{TT}
\end{figure}

\begin{figure}
  \includegraphics[width=12cm]{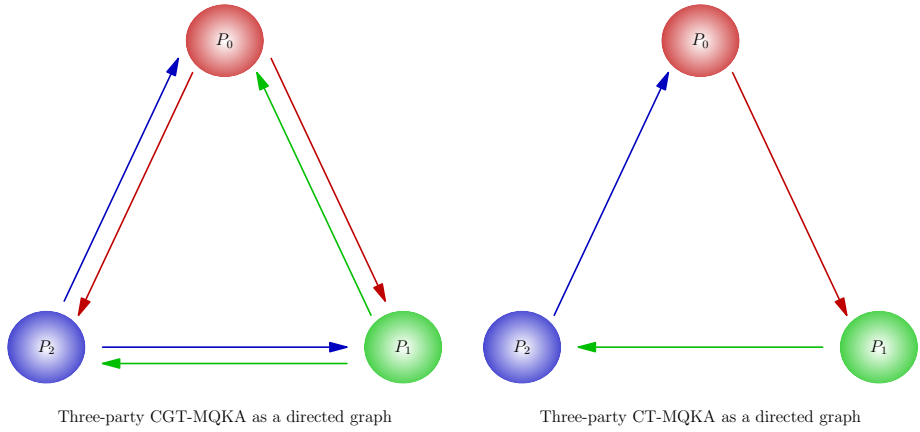}\\
  \caption{The protocols in \cite{MQKA_Yin,He15} can be considered as the left directed graph, while the protocols in \cite{MQKA_Yin2,Chitra14,Zhu15} can be considered as the right one.}\label{3party}
\end{figure}

If we consider the participants and the transmissions of quantum states between them as the vertices and the edges in a graph respectively, the protocols \cite{QKA_Liu,MQKA_Yin,He15} of the first kind are complete graphs, and the protocols \cite{QKA2012e,MQKA_Yin2,SunvsLiu,Chitra14,Zhu15,Sun2015QIP,Sun2015IJTP} of the second kind are circles (See Fig. \ref{TT}).
And this is why we call them the complete-graph-type MQKA and circle-type MQKA respectively.
For the three-party ones \cite{MQKA_Yin,He15,MQKA_Yin2,Chitra14,Zhu15}, where the complete graphs are just the circles, we classify them by considering their extensional versions to $N$-party ones, where $N$$>$3. Or, we can classify them by simply checking whether they are directed complete graphs, for example, protocols in \cite{MQKA_Yin,He15} are, but ones in \cite{MQKA_Yin2,Chitra14,Zhu15} are not (See Fig. \ref{3party}).
The protocol in Ref. \cite{Xu14} is a tree graph with only one root (Alice). Therefore, the protocol in Ref. \cite{Xu14} is called the tree-type MQKA.
In fact, the tree-type QKA protocol is sensitive to a special attack called detection bits chosen attack \cite{QKA_IF} which we will briefly introduce in conclusions.

%While, when classical communications are also considered, this protocol becomes a complete graph, so we also classify into the complete-graph-type MQKA.

%The other is that the protocols in Ref. \cite{QKA2012e} and the improved version \cite{Zhu15} for the protocol in Ref. \cite{Chitra14} are circle when only the quantum transmissions are considered as the edges. Here we classify them in circle-type MQKA because the classical communications

Obviously, CT-MQKA protocols need less communications and less quantum channels than CGT-MQKA ones for the same number of participants. This is the main reason why people tend to design circle-type ones. However, CT-MQKA protocols face more threatens on fairness than CGT-MQKA protocols. in next section, we will introduce the collusive attacks against CT-MQKA protocols.

Above categories only considered the transmission of quantum particles. If the classical communications are also considered, the protocols in Refs. \cite{Xu14,QKA2012e,Zhu15} become complete graphes, and we call them classical complete-graph-type MQKA protocols.

\section{Collusive Attacks Against Circle-Type MQKA Protocols}

%The attack strategies which aims to predetermine the final key in QKA are similar in the sense that they are all participant attacks where the dishonest participants manage to get the legal final key before the protocol has finished and then try to flip parts of its bits according to their expected one in the rest processes of the protocol. The attack strategy we will describe next is in the same manner.

We first formalize the multi-party CT-MQKA protocols that we are attacking. Suppose there are $N$ participants $P_0$, $P_1$, $\ldots$, $P_{N-1}$ and their personal keys are $K_0$, $K_1$, $\ldots$, $K_{N-1}$, respectively.

At the beginning of the protocol, $P_i$ generates a sequence of entangled states $|\Psi_i\rangle$ \footnote{The single states generated in some protocols can be considered as the entangled states where parts of them ($R_i$) have already been measured, just like that in the security proof of BB84 \cite{BB84proof}.} and divides each state into two parts, one of which will be kept in his/her hand and the other will be sent out. And we denote the sequence of the two parts as $R_i$ and $S_i$, respectively, where $i$$=$0, 1, $\ldots$, $N$$-$1.

Then all the $S_i$s are transmitted in the same direction in the circle. Once all the $S_i$s have been transmitted from one participant to the next one, the participants keep the sequences that they have just received for a while, during which they perform the detection and encode their personal keys in the received sequences. Afterwards, they continue to send the above sequences to the next participant.

After passing through all the other participants, each sequence would be sent back to the participant who generated it (finishing a complete circle). Then $P_i$ can measure $R_i$ and $S_i$ to get the bitwise exclusive OR results of all the other participants' personal keys. Finally, they can calculate the final key $K_\textrm{final}$$=$$\bigoplus^{N-1}_{i=0} K_i$.

For the convenience of description, we divide the whole process of the above $N$-party protocol into $N$ periods.

In the first period, $P_i$ generates $S_i$ and $R_i$ and send $S_i$ \footnote{In fact, $S_i$ has changed since $P_i$ has probably inserted some decoy particles in it. And later, other participant will encode their secrets in it and also insert their decoy states in it. However, for simplicity, we call all the sequences which include the particles of $S_i$ simply $S_i$.} to $P_{i\boxplus 1}$, where ``$\boxplus$'' represents addition modular $N$ and ``$\boxminus$'' below represents subtraction modular $N$.

The $k$-th ($2$$\leq$$k$$\leq$$N$$-$1) period starts from the moment when each participant $P_i$ has received the sequence $S_{i\boxminus (k-1)}$, which is prepared by $P_{i\boxminus (k-1)}$ in the first period and sent from $P_{i\boxminus1}$ in the $(k$$-$$1)$-th period.
And in the $k$-th period, the participant $P_i$ performs the detection processes \footnote{The detection processes generally contains three stages, publishing the positions of the decoy states, measuring them, and comparing the results.} with $P_{i\boxminus1}$ and $P_{i\boxplus1}$ to detect the possible attacks on $S_{i\boxminus (k-1)}$ and $S_{i\boxminus k}$ in the $(k$$-$$1)$-th period respectively, and then encodes his/her personal key $K_i$ in $S_{i\boxminus (k-1)}$. Then $P_i$ inserts some decoy states in it and sends it to $P_{i\boxplus1}$. The $k$-th period ends and the $(k$$+$$1)$-th period starts when all the participants have received the sequences sent to them in the $k$-th period.

In the last (the $N$-th) period, $P_i$ performs the detection processes with $P_{i\boxminus1}$ and $P_{i\boxplus1}$ as before. Then $P_i$ measures $R_i$ and $S_i$ to get the bitwise exclusive OR result of the others' personal keys. Finally, $P_i$ obtains the final key by performing a bitwise exclusive OR on the above result and $K_i$.

All the CT-MQKA protocols can be described and disassembled as above, except the one in Ref. \cite{Sun2015IJTP}. Nevertheless, the collusive attack which we will introduce next can also attack it successfully.
Now we introduce the collusive attacks against CT-MQKA protocols, which can be divided into two stages: key stealing stage and key flipping stage. In key stealing stage, the dishonest participants manage to get the bitwise exclusive OR result of the others' personal keys. And in the key flipping stage, they flip the encoded personal keys according to the above result to control the final key.

\begin{figure}
  \includegraphics[width=12cm]{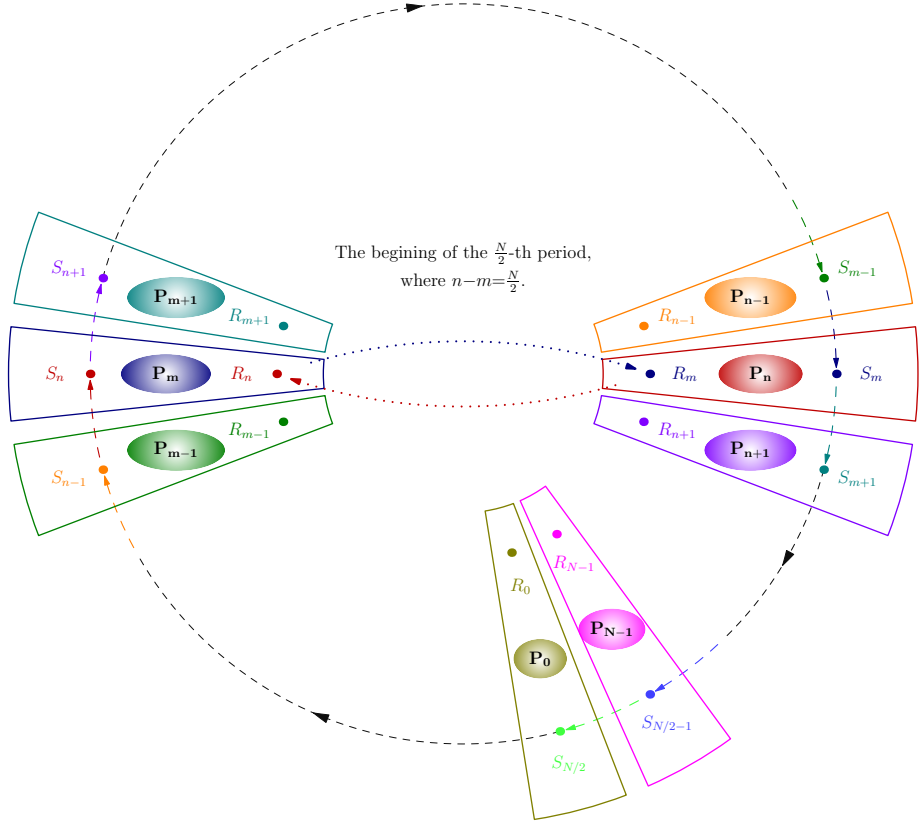}\\
  \caption{The collusive attacks to CT-MQKA protocols in the $(N/2)$-th period, where $N$ is an even number and $n$$-$$m$$=$$N/2$. The areas circled by the color lines belong to the participants with the same color. The dashed lines and the dotted lines represent the transmission of quantum particles happened in last (the ($N/2$$-$1)-th) period and the first period, respectively. In the $(N/2)$-th period, $P_n$ can measure $S_m$, which is sent by $P_{n-1}$ in last period, and $R_m$, which is sent by $P_m$ in the first period, to get the bitwise exclusive OR result of $K_{m+1}$, $K_{m+2}$, $\ldots$, $K_{n-1}$. Similarly, $P_n$ can get the bitwise exclusive OR result of $K_{n+1}$, $K_{n+2}$, $\ldots$, $K_{N-1}$, $K_0$, $\ldots$, $K_{m-1}$}\label{attack}
\end{figure}

In fact, any two participants $P_n$ and $P_m$ ($n$$>$$m$) can cooperate to get the bitwise exclusive OR results of the personal keys belong to the participants between them at a certain period (the $(n$$-$$m)$-th and the $(N$$-$$n$$+$$m)$-th). The attack process of the key stealing stage can be described as follows.
\begin{itemize}
  \item [1] In the first period, $P_n$ ($P_m$) sends the information about what the states he/she prepared initially and the sequence $R_n$ ($R_m$) which should be kept in his/her hand to $P_m$ ($P_n$). This is equivalent to switching the positions of each other. Of course, they also share the value of excepted key which they are trying to make the final key be.
  \item [2] In the $(n$$-$$m)$-th  period, $P_n$  has received the sequence $S_m$, which carries the bitwise exclusive OR result of the personal keys $K_{m+1}$, $K_{m+2}$, $\ldots$, $K_{n-1}$. Since $P_n$ owns $R_m$, $S_m$ and the initial state of them at present, he/she can extract the above bitwise exclusive OR result in this period just like what $P_m$ should do in the last period. Similarly, $P_m$ can get the bitwise exclusive OR result of the personal keys $K_{n+1}$, $K_{n+2}$, $\ldots$, $K_{N-1}$, $K_{0}$, $\ldots$, $K_{m-1}$ in the $(N$$-$$n$$+$$m)$-th period. (See Fig. \ref{attack})
  \item [3] $P_n$ ($P_m$) sends the above bitwise exclusive OR result to $P_m$ ($P_n$) immediately he/she gets it.
\end{itemize}

In the collusive attacks to CT-MQKA protocols, timeliness is a very important factor. Because of the decoy states, the dishonest participants cannot tamper (flip) the personal keys encoded by other participants without being detected. Therefore, to successfully control the final key, the dishonest participants should ensure that all the sequences prepared by the other participants will pass through them at least once after they have get the bitwise exclusive OR result of all the others' personal keys. Actually, two dishonest participants are enough to totally control the final key, as long as their positions in the circle satisfy the following conditions,
\begin{eqnarray}
  n-m=\frac{N}{2} &\ \ & \textrm{for an even } N; \label{even} \\
  n-m=\frac{N-1}{2} \textrm{\ or\ } \frac{N+1}{2} &\ \ & \textrm{for an odd } N. \label{odd}
\end{eqnarray}
We first consider the situation when $N$ is an even number. In the $(N/2)$-th period, each of $P_n$ and $P_m$ gets the bitwise exclusive OR result of half of the others' personal keys. After exchanging with each other, they can know what the legal final key $K_\textrm{final}$ should be. Then $P_n$ and $P_m$ will encode
\begin{equation}\label{1}
    K^\prime_n=K_n\oplus K_\textrm{expected}\oplus K_\textrm{final}
\end{equation}
instead of $K_n$ and
\begin{equation}\label{2}
    K^\prime_m=K_m\oplus K_\textrm{expected}\oplus K_\textrm{final}
\end{equation}
instead of $K_m$ respectively in the rest periods, where $K_\textrm{expected}$ is what they want the final key to be. Then $P_n$ can flip the sequences $S_{m-1}$, $S_{m-2}$, $\ldots$, $S_{0}$, $S_{N-1}$, $\ldots$, $S_{n+1}$ and $P_m$ can flip the sequences $S_{n-1}$, $S_{n-2}$, $\ldots$, $S_{m+1}$. Thus, in the last period, for any participant $P_i$, he/she will get the
\begin{eqnarray}\label{3}
    K^\prime_\textrm{final}&=&K_0\oplus K_1\oplus\ldots\oplus(K_{n\textrm{ or }m}\oplus K_\textrm{expected}\oplus K_\textrm{final})\oplus\ldots\oplus K_{N-1}\nonumber\\
    &=&K_\textrm{final}\oplus K_\textrm{expected} \oplus K_\textrm{final}\\
    &=&K_\textrm{expected}.\nonumber
\end{eqnarray}

For the odd $N$, the situation is similar. In the $((N+1)/2)$-th period, one of the dishonest participants will get the bitwise exclusive OR result of $(N-1)/2$ personal keys, while the other has already got the result of rest $(N-3)/2$ personal keys in last period. Just like what they do when $N$ is even, the former dishonest participant, for example $P_n$, can flip the sequences $S_{m-1}$, $S_{m-2}$, $\ldots$, $S_{0}$, $S_{N-1}$, $\ldots$, $S_{n+1}$, and the latter $P_m$ can flip the sequences $S_{n-1}$, $S_{n-2}$, $\ldots$, $S_{m+1}$. Note that the difference between the two cases is that $P_m$ should flip $S_{n-1}$ in the period when he/she knows the final key (in the $((N+1)/2)$-th period) when $N$ is odd, and in the next period (in the $(N/2+1)$-th period) when $N$ is even.

In fact, as long as the longest distance in the circle between adjacent dishonest participants is no more than $\lfloor(N+1)/2\rfloor$, they can successfully control the final key, where $\lfloor x\rfloor$ represents the maximum integer which is not more than $x$.
In the $\lfloor(N+1)/2\rfloor$-th period of this situation, the dishonest participants can always get the legal final key during the past and on-going periods [0,$\lfloor(N+1)/2\rfloor$]. And the length of the on-going period and the periods that have not yet passed [$\lfloor(N+1)/2\rfloor$, $N$] is not shorter than that of [0,$\lfloor(N+1)/2\rfloor$]. Therefore, in the periods [$\lfloor(N+1)/2\rfloor$, $N$], they always have an opportunity to flip all the $S_i$s.

For example, three dishonest participants $P_0$, $P_{N/3}$ and $P_{2N/3}$ can also succeed although no two of them satisfy Eq. \ref{even} or Eq. \ref{odd}. Concretely, at the $(N/3)$-th period, they can get the legal final key, and in the next $N/3$ periods, they remain to perform the legal operations, while in the last $N/3$ periods, they flip the received sequences by encoding the tampered key as $K^\prime_n$ and $K^\prime_m$ above.

All the CT-MQKA protocols are sensitive to the above collusive attacks, although the situation for the one in Ref. \cite{Sun2015IJTP} is a little different. In this specific protocol, each participant sends out two sequences, each of which ``runs'' half circle. For the collusive attack against this protocol, two dishonest participants cannot succeed any more, since they can only get half of the others' personal keys before the last period, which leaves no time for them to flip the others' sequences. For this specific protocol, three dishonest participants are necessary and they can succeed as long as they do not located in a same minor arc of the circle, i.e., an arc whose length is less than half of the circumference.

In fact, the attack strategy described above is an instructional mode of the attacks to CT-MQKA protocols but not a detailed attack to any specific protocol. In the attacks to specific MQKA protocols, just as what we have stated above, the manner for the dishonest participants to steal the personal keys of the others is the same with the manner for the honest participants to extract the bitwise exclusive OR result, and the way for the dishonest participants to flip the legal key is the same with the way for the honest participants to encode their personal keys. In fact, some attack strategies proposed before are similar to the one we proposed here; in other words, the proposed model is a summary of the previous attacks.

For example in the attack strategy proposed by Huang et al. \cite{HuangvsSun}, $N$$-$1 dishonest participants try to predetermine the final key. The key stealing stage is performed by $P_0$ and $P_{N-2}$, they can steal $P_{N-1}$'s personal key by performing the measurement on $Z$ basis or $X$ basis (according to the initial bases of $S_{N-2}$) on each particle in $S_{N-2}$. The key flipping stage is performed by $P_{N-2}$, where he flips $P_{N-1}$'s measurement results by performing $I$ or $iY$ on each particle in $S_{N-1}$.

In Zhu et al.'s attack strategy \cite{Zhu15} to the three-party MQKA protocol in Ref. \cite{Chitra14}, two dishonest participants (Alice and Bob) try to cheat the honest one (Charlie). In Step 5 of the original protocol, Alice and Bob can deduce Charlie's encoding operations by performing Bell measurement to the particle pairs in the two sequences $p_B$ and $r_C$, where the corresponding particles in $p_B$ and $r_C$ are initially generated in a Bell state by Bob, and $r_C$ has been operated by Charlie to encode his personal key $K_C$. The above attack actions are corresponding to the key stealing stage in our attack mode. Then in Step 6, Alice and Bob performs $I$ or $Z$ according to Charlie's operations and their expected key to control the final key. And the above actions are corresponding to the key flipping stage in our attack mode.
However, in their improvement \cite{Zhu15} of the protocol in Ref. \cite{Chitra14}, each of the three participants encrypts his/her personal key with another key, and before the detection in the last (third) period, they announces the latter keys, denoted as additional keys. The authors \cite{Zhu15} think the improved version is fair. However, two dishonest participants, for example Alice and Bob, can still predetermine the final key. By the collusive attacks introduced above, Alice and Bob knows what Charlie, the honest one, has encoded in their sequences, so they know what the bitwise exclusive OR result of the three keys encoded in the sequences. Thus, they can find a way to get Charlie's additional key earlier, then announce the false additional keys to control the final key. This loophole in this improved version is similar with that in \cite{QKA2012e}.

\section{Discussions}

In this paper, we introduce the collusive attacks to CT-MQKA protocols. Research shows that all the CT-MQKA protocols are unfair against collusive attacks. Here we summarize the fairness of all the existing MQKA protocols in Table 1.
\begin{table}
\centering
\caption{The fairness of all the existing MQKA protocols. Here, CGT represents that the protocol is a complete-graph-type MQKA protocol, TT represents tree-type one, CT represents circle-type one, and CCGT represents that this protocol is also a classical complete-graph-type one.}
%\label{tab2}\vspace{1mm}
%\renewcommand{\arraystretch}{1.15}
\begin{tabular}{c|c|c|c|c}
%\multicolumn{2}{ c }{ }\\
  \multirow{2}*{Protocol} & \multirow{2}*{Category} &Fair against & Fair against & \multirow{2}*{Comments}\\
                        & &                         single attacks? & collusive attacks?& \\
  \hline\hline
  %\\
  \cite{QKA_Liu}   & CGT & Yes & Yes& \\\hline
  \cite{MQKA_Yin}  & CGT & Yes & Yes& \\\hline
  \cite{He15}      & CGT & Yes & Yes&\\\hline
  \multirow{3}*{\cite{Xu14}}      & \multirow{3}*{TT/CCGT}  & \multirow{3}*{No}  & \multirow{3}*{No} & The attack effect depends\\
                                &&&&     on the proportion of the\\
                                &&&&     detection states \cite{QKA_IF}\\\hline
  \cite{QKA2012e}  & CT/CCGT & No  & No & Has been analyzed in \cite{QKA_Liu}\\\hline
  \cite{MQKA_Yin2} & CT  & Yes & No &\\\hline
  \cite{SunvsLiu}  & CT  & Yes & No & Has been analyzed in \cite{HuangvsSun}\\\hline
  \cite{Chitra14}  &  CT & Yes & No & Has been analyzed in \cite{Zhu15}\\\hline
  \cite{Zhu15}     &  CT/CCGT & Yes & No &\\\hline
  \cite{Sun2015QIP}&CT   & Yes & No &\\\hline
  \cite{Sun2015IJTP}&CT  & Yes & No & \\\hline
\end{tabular}
\end{table}

In fact, the tree-type MQKA protocol \cite{Xu14} is unfair since the participants who perform the detection process later can control the key to some extent. Considering the following simple example of the protocol in Ref. \cite{Xu14} where three participants, Alice (who generates the GHZ states), Bob and Charlie, are generating a 2-bit key with 5 GHZ states, and each of them chooses one state to detect the attacks. In this case, if Bob first chooses one state randomly to perform his detection, Alice and Charlie might (partly) predetermine the key through the following detections. Once Bob has finished his detection, Alice measures all the rest 4 GHZ states and chooses her and Charlie's detection states according to the measurement results and their expected key. For example, the measurement results are 0101 and their expected key is 10. Thus they could choose the first and the last states and pretend to perform detections as usual. The final key would be 10. Peculiarly, according to the results in Ref. \cite{QKA_IF}, if the number of dishonest participants' detect bits is larger than that of the final key, they can always totally predetermine the key.

Is it possible to design a fair CT-MQKA protocol? What is clear is that it is impossible for a ``pure'' CT-MQKA protocol, according to the analysis above. Here by ``pure'' CT-MQKA protocols we mean the MQKA protocols which are CT-MQKA protocols and are not classical complete-graph-type MQKA protocols. And whether the complete-graph-type classical communications can solve the fairness problem of CT-MQKA is an open question.

\begin{acknowledgements}
This work is supported by the National Natural Science Foundation of China under Grant Nos. 61572089, 61309029, 61502200, the Natural Science Foundation of Guangdong Province under Grant No.2014A030310245, the Fundamental Research Funds for the Central Universities under Grant Nos. 0903005203369, 21615313.
\end{acknowledgements}



\end{document}